\documentclass[1p]{elsarticle}

\usepackage{graphicx}
\usepackage{hyperref}
\usepackage{amssymb}
\usepackage{amsmath}
\usepackage{amsfonts}
\usepackage{epstopdf}
\def\simleq{\; \raise0.3ex\hbox{$<$\kern-0.75em \raise-1.1ex\hbox{$\sim$}}\; }
\def\simgeq{\; \raise0.3ex\hbox{$>$\kern-0.75em \raise-1.1ex\hbox{$\sim$}}\; }

\newcommand{\GeV}{{\rm GeV}}
\newcommand{\TeV}{{\rm TeV}}

\newcommand{\erg}{{\rm erg}}

\newcommand{\kpc}{{\rm kpc}}
\newcommand{\pc}{{\rm pc}}

\newcommand{\cm}{{\rm cm}}

\newcommand{\s}{{\rm s}}
\newcommand{\sr}{{\rm sr}}

\newcommand{\fraction}[2]{\frac{\displaystyle #1}{\displaystyle
#2}}

\begin{document}
\title{On possible interpretations of the high energy electron-positron spectrum measured by the Fermi Large Area Telescope}

\author[infn_pisa]{D.~Grasso \corref{cor1}}
\ead{dario.grasso@pi.infn.it}

\author[santa_cruz]{S.~Profumo\corref{cor1}}
\ead{profumo@scipp.ucsc.edu}

\author[mpe]{A.W.~Strong \corref{cor1}}
\ead{aws@mpe.mpg.de}

\author[infn_pisa]{L.~Baldini}

\author[infn_pisa]{R.~Bellazzini}

\author[slac]{E.~D.~Bloom}

\author[infn_pisa]{J.~Bregeon}

\author[infn_pisa,unipisa]{G.~Di Bernardo}

\author[infn_pisa,unipisa]{D.~Gaggero}

\author[unibari,infn_bari]{N.~Giglietto}

\author[slac]{T.~Kamae}

\author[infn_pisa]{ L.~Latronico}

\author[infn_trieste,unitrieste]{F.~Longo}

\author[infn_bari]{M.N.~Mazziotta}

\author[cresst,unimaryland]{A.~A.~Moiseev}

\author[infn_romatv]{A.~Morselli}

\author[unidenver]{J.F.~Ormes}

\author[infn_pisa]{M.~Pesce-Rollins}

\author[uniiowa]{M.~Pohl}

\author[infn_pisa]{M.~Razzano}

\author[infn_pisa]{C.~Sgro}

\author[infn_pisa]{G.~Spandre}

\author[nasa]{T.~E.~Stephens}
\noindent
\cortext[cor1]{Corresponding author}

\address[infn_pisa]{Istituto Nazionale di Fisica Nucleare, Sezione di Pisa,  I-56127 Pisa, Italy}
\address[santa_cruz]{Santa Cruz Institute for Particle Physics and Department of Physics,
University of California, Santa Cruz, CA 95064, USA}
\address[mpe]{ Max-Planck Institut f\"ur extraterrestrische Physik, 85748 Garching, Germany}
\address[slac]{Kavli Institute for Particle Astrophysics and Cosmology,
Department of Physics and SLAC National Accelerator Laboratory, Stanford University, Stanford, CA 94305, USA}
\address[unipisa]{Dipartimento di Fisica, Universit\`a di Pisa, I-56127 Pisa, Italy}
\address[unibari]{Dipartimento di Fisica ÒM. MerlinÓ dell' Universit\`a e del Politecnico di Bari, I-70126 Bari, Italy}
\address[infn_bari]{Istituto Nazionale di Fisica Nucleare, Sezione di Bari, 70126 Bari, Italy}
\address[infn_trieste]{Dipartimento di Fisica, Universit\`a di Trieste, I-34127 Trieste, Italy}
\address[unitrieste]{Dipartimento di Fisica, Universit\`a di Trieste, I-34127 Trieste, Italy}
\address[cresst]{Center for Research and Exploration in Space Science and Technology (CRESST), NASA Goddard Space Flight Center, Greenbelt, MD 20771, USA}
\address[unimaryland]{University of Maryland, College Park, MD 20742, USA}
\address[infn_romatv]{Istituto Nazionale di Fisica Nucleare, Sezione di Roma ÒTor VergataÓ, I-00133 Roma, Italy}
\address[unidenver]{Department of Physics and Astronomy, University of Denver, Denver, CO 80208, USA}
\address[uniiowa]{Iowa State University, Department of Physics and Astronomy, Ames, IA 50011-3160, USA}
\address[nasa]{Universities Space Research Association, NASA Ames Research Center, MS211-3 Moffet Field, CA 94035}

\begin{abstract}
The Fermi-LAT experiment recently reported high precision measurements of the spectrum of cosmic-ray electrons-plus-positrons (CRE) between 20 GeV and 1 TeV. 
The spectrum  shows no prominent spectral features, and is significantly harder than that inferred from several previous experiments. 
Here we discuss several interpretations of the Fermi results based either on a single large scale Galactic CRE component or by invoking additional 
electron-positron primary sources, e.g. nearby pulsars or particle Dark Matter annihilation.  We show that while the reported Fermi-LAT data alone can be interpreted in terms of a single component scenario, when combined with other complementary experimental results, specifically the CRE spectrum measured by H.E.S.S. and especially the positron fraction reported by PAMELA between 1 and 100 GeV,  that class of models fails to provide a consistent interpretation. Rather, we find that several combinations of parameters, involving both the pulsar and dark matter scenarios, allow a consistent description of those results. 
We also briefly discuss the possibility of discriminating between the pulsar and dark matter interpretations by looking for a possible anisotropy in the CRE flux. 
\end{abstract}
\maketitle

\section{Introduction}

Measuring the spectrum of Cosmic Ray  electrons (CRE) (unless explicitly otherwise stated we define electrons to be electrons + positrons for this paper) with high accuracy and over a wide energy range is important to constrain theoretical models of production and propagation of CRs in the Galaxy, including signatures of new physics. So far, direct measurements extend over more than six decades of energy, from MeV to some TeV. While at low energy (up to few GeV) solar modulation plays an important role in determining the spectral shape, at higher energies the spectrum is expected to be determined mainly by three elements: the slope of the source injection power-law, synchrotron and Inverse Compton (IC) energy losses, and diffusion in the turbulent Galactic magnetic fields. For this reason the high-energy part of the spectrum is the most interesting when trying to constrain theoretical models.

Prior to 2008, the high energy electron spectrum was measured by balloon-borne experiments (Kohayashi et al. 2004 ~\cite{Kobayashi:2003kp}) and by a single space mission AMS-01 (Aguilar et al. 2002 ~\cite{ams1}). Those data are compatible with a featureless power-law spectrum within their errors. This is in agreement with theoretical predictions from both analytical and numerical calculations (for a recent review see Strong et al. 2007 ~\cite{Strong:2007nh}) in which: i) the source term of CR electrons is
treated as a time-independent and smooth function of the position in the Galaxy, and the energy dependence is assumed to be a power law; ii) the propagation is described by a diffusion-loss equation whose effect is to steepen the spectral slope with respect to the injection. 

It is important to remember that astrophysical sources of electrons are actually stochastic in space and time. Since electron propagation - in contrast with the hadronic part of CRs - is severely limited by energy losses via synchrotron radiation and IC scattering, a large fraction of the electrons that are detected above 100 GeV are expected to be produced within few kpc of the Earth by few sources. Statistical fluctuations in the injection spectrum and spatial distribution of those nearby sources may produce significant deviations in 
the most energetic part of the observed spectrum compared to the conventional homogeneous and steady state scenario (see e.g. Atoyan et al. 1995 \cite{Atoyan:95},
Pohl  \& Esposito 1998 ~ \cite{Pohl:1998ug}, Strong \& Moskalenko 2001 ~\cite{Strong:2001qp}, Kobayashi et al. 2004 ~\cite{Kobayashi:2003kp}).

Recently, the ATIC balloon experiment (Chang et al. 2008 ~\cite{atic}) found a prominent spectral feature 
at around 600 GeV in the total electron spectrum \footnote{Fazely et al. (2009~\cite{Fazely:2009jb}) claimed, however, that ATIC excess may be interpreted as a contribution
of misidentified protons.}.  This feature was also marginally observed by PPB-BETS (Tori et al. 2008 ~\cite{Torii:2008xu}). Furthermore,
the H.E.S.S. (Aharonian et al. 2008, 2009~\cite{hess,hess:09}) atmospheric Cherenkov telescope (ACT) reported a significant steepening of the electron  spectrum above $\sim 1$ TeV.

In addition to (charge undifferentiated) electron measurements, another independent indication of the presence of a possible deviation from the standard picture came from the recent measurements of 
the positron to electron fraction, e$^+$ /(e$^-$+ e$^+$), between 1.5 and 100 GeV by the PAMELA satellite experiment (Adriani et al. 2009, 2009b~\cite{PAMELA,PAMELA_Nature}).  
PAMELA found that the positron fraction changes slope at around 10 GeV and begins to increase steadily up to 100 GeV. A similar trend was also indicated by earlier
experiments  HEAT (Barwick et al. 1997 ~\cite{Barwick:1997ig}) and AMS-01 (Aguilar et al. 2002 ~ \cite{ams1}) (see also Bellotti et al. \cite{Wizard})
with lower significance and in a narrower energy range. 
This behavior is very different from that predicted for secondary positrons produced in the collision of CR nuclides with the interstellar medium (ISM)  (see e.g. Moskalenko \& Strong 1998a).
The discrepancy moderates only if one considers a very steep injection index for electrons (Delahaye et al. 2008 ~\cite{Delahaye:2008ua}).

Based on their observations, the recent publications of the ATIC (Chang et al. 2008 ~\cite{atic}) and PAMELA (Adriani et al. 2009b~\cite{PAMELA_Nature}) collaborations report the need for an additional component of electrons and positrons originating from pulsars, or dark matter, clearly unaccounted for in the standard CR model. Indeed, the possibility that the excess of high-energy positrons measured by PAMELA and the anomalous spectrum reported by ATIC and PPB-BETS in the several hundreds of GeV range are connected with a dark matter particle has stirred great interest 
(for early references see e.g.~\cite{Cirelli:2008pk,Bringmann:2009ip,ArkaniHamed:2008qn,Fox:2008kb,Harnik:2008uu}). Astrophysical interpretations of PAMELA results, based on the role of one (or more) nearby pulsars (see e.g. Hooper at al. 2008 ~\cite{Hooper:2008kg}, Yuksel et al. 2008 ~\cite{Yuksel:2008rf})
have also been proposed although a combined interpretation of ATIC and PAMELA results in that framework was shown to be unlikely (Profumo, 2008 ~\cite{Profumo:2008ms}). 

Very recently the experimental information available on the CRE spectrum has been dramatically expanded as the Fermi-LAT Collaboration has reported a high precision measurement of the electron spectrum from 20 GeV to 1 TeV performed with its Large Area Telescope (LAT) 
(Abdo et al. 2009\  \cite{Fermi_CRE_1})).  As Fig. \ref{fig:elepos_242reac} shows, a simple power-law fit of the Fermi-LAT electron energy spectrum is possible giving: 
\begin{eqnarray}
J_{e^\pm} = (175.40 \pm 6.09) \left(\frac{E}{1~\GeV} \right)^{-(3.045 \pm 0.008)}  \GeV^{-1} {\rm m}^{-2} {\rm s}^{-1} {\rm sr}^{-1}                         
\label{eqn:eqn1}
\end{eqnarray}
with $\chi^2$ = 9.7 (for 23 d.o.f.) where statistical and systematic (dominant) errors have been, conservatively, added in quadrature. 
The systematic error on the Fermi-LAT energy calibration may also result in a +10\%, - 20\% rigid shift of the spectrum without introducing signiÞcant deformations. 
Again referring to Fig. \ref{fig:elepos_242reac}, this spectrum agrees with ATIC below 300 GeV, but Fermi-LAT does not confirm the prominent spectral feature observed by ATIC at larger energies. 
Very recently the H.E.S.S. collaboration released a new set of data for the CRE electron spectrum in the $340~ \GeV - 5~\TeV$ energy range. Those data agree with Fermi-LAT's, within their systematic errors, in the energy range covered by both
experiments while at larger energies H.E.S.S. report a significant spectral steeping (Aharonian et al. 2009~\cite{hess:09}).  

Looking almost featureless at first glance, the electron spectrum measured by Fermi-LAT reveals a hardening at around 100 GeV and a steepening above $\sim 400$ GeV. 
Indeed, 
the spectrum can be fitted by a broken power-law with indexes  $- 3.070 \pm 0.025$ for $E  < 100~\GeV$,  $-2.986 \pm 0.031$ for $100 < E < 400~\GeV$ and 
 $- 3.266 \pm 0.116$ for $400 < E  < 1000~\GeV$. 
While we cannot claim any deviation from a single power-law when conservatively taking into account current systematic uncertainties, such features are suggestive
 when trying to combine Fermi, H.E.S.S., PAMELA and low-energy electron data for various interpretations. It is worth noticing here that, although Fermi results damp some of the expectations excited by the ATIC results, the hard electron spectrum observed by this experiment exacerbates the discrepancy between the predictions of standard CR theoretical models and the positron fraction excess measured, most conclusively, by PAMELA. This makes the exploration of some non-standard interpretations more compelling. 

In this paper we consider several interpretation scenarios for the CRE data reported by Fermi-LAT. In Sec.\ref{sec:GCRE}, we start by considering a conventional
{\it Galactic CR electron scenario} 
(GCRE) model assuming that electrons are accelerated only by continuously distributed astrophysical sources (probably Supernova Remnants (SNR)) in the Galactic disk, plus a secondary component of electrons and positrons produced by the collision of primary CR nuclides with the interstellar gas.
In Sec. \ref{sec:pulsars} we account for the contribution of nearby, observed astrophysical sources. We focus in particular on pulsars, since these objects are undisputed sources
of electron and positron pairs offering a natural interpretation not only to the Fermi and H.E.S.S. CRE data but also to the PAMELA measurement of the positron fraction.  
Dark matter (DM) annihilation also offers a viable scenario to interpret the current CRE experimental results. In Sec. \ref{sec:DM} we carry out a study of prototypical classes of particle DM models, and we study the relevance of Fermi-LAT CRE data in constraining the model parameter space.     
For both the pulsar and the DM scenarios we also briefly discuss the consistency of the proposed models and the possibility of testing them with current and future gamma-ray measurements by Fermi. 
In the discussion section \ref{sec:discussion} we consider possible signatures which may allow disentangling the different interpretations of the Fermi-LAT CRE
results discussed in this paper.

\section{Interpreting Fermi data with a large-scale Galactic CRE component}\label{sec:GCRE}

\subsection{The case a smooth Galactic CRE component}\label{sec:galprop3}

In this section we try to interpret Fermi-LAT, H.E.S.S., and low energy electron data in terms of a  conventional CR diffusion model using the GALPROP package when appropriate. The same package will be used in other parts of this work. Where necessary other methods are used, for example for modeling nearby pulsar sources or DM sources. The numerical CR propagation code GALPROP (Moskalenko and Strong, 2001 ~\cite{Moskalenko:01}) is designed to make predictions of many kinds of observational data: CR direct measurements including primary and secondary nuclei, electrons and positrons, gamma rays, synchrotron radiation. After the CR propagation has been computed for all species including secondaries, the CR spectrum at each point in the Galaxy is used to compute gamma-rays using gas surveys and a detailed model of the interstellar radiation field (Porter et al. 2005 ~\cite{Porter:2008ve}). Synchrotron radiation is computed using the electron and positron spectra and a 3-dimensional model of the Galactic magnetic field. For the application to the Fermi electron measurements, it is an advantage that GALPROP is also used for the Fermi gamma-ray predictions, furthering a consistent approach 
\footnote{GALPROP is a public code but is in continuous development by a small team, and the current version v.54 is used here.  A detailed description can be found at
http://galprop.stanford.edu.}. 

The main parameters for a given GALPROP model are the CR primary injection spectra, the spatial distribution of CR sources, the size of the propagation region, the spatial and momentum diffusion coefficients and their dependence on particle rigidity. The propagation parameters have been chosen to fit the boron to carbon
(B/C) ratio, radioactive nuclei and the Galactic distribution of CR sources from previous gamma ray studies (Strong et al. 2004 ~\cite{Strong:04}).  
The only adjustment to Fermi-LAT CRE data is for the electron injection spectrum. Tab. \ref{table1} summarizes the main parameters used in this paper.
The low energy index is chosen to avoid overproducing gamma-rays at low energies seen by other experiments  (see Strong et al. 2004 ~\cite{Strong:04} for details). 
 All models considered here are based on the locally observed electron and nucleon spectra. Following the notation generally adopted in the literature,  we name those model as ``conventional models"  to distinguish them from ``optimized models" which assume modified local spectra (see Strong et al. 2004 ~\cite{Strong:04}).

\begin{table}[ht]
\centering
\caption{Propagation and CR injection parameters for the GCRE models considered in this paper. $D_0$ is the diffusion coefficient normalization at $1~\GeV$; 
$\delta$ the index of the power-law dependence of $D$ on energy; $z_h$ the half-width of the Galactic CR confinement halo;  $\gamma_0$ the electron 
injection power-law index;  $N_{e^-}$ is the  electron flux normalization at $E = 100~\GeV$;   $\gamma^p_0$ the CR nuclei injection index.
Models 0 and 1 account for CR re-acceleration in the ISM, while 2 is a plain-diffusion model. 
}
\begin{center}
\begin{tabular}{|c|c|c|c|c|c|c|}
\hline
\hline 
Model \#& $D_0\  (cm^2\s^{-1})$  & $\delta$ &  $z_h\ (\kpc)$ & $\gamma_0$ &  $N_{e^-}\ (m^{-2} s^{-1} \sr^{-1} \GeV^{-1})$ & $\gamma^p_0$ \cr
\hline
 0  &   $3.6 \times 10^{28}$ & 0.33 & 4 & 2.54 & $1.3\times10^{-4}$ & 2.42 \cr
 1  &   $3.6 \times 10^{28}$ & 0.33 & 4 & 2.42 & $1.3\times10^{-4}$ & 2.42 \cr
 2  &    $1.3 \times 10^{28}$ & 0.60 & 4 & 2.33 &  $1.3\times10^{-4}$ & 2.1 \cr
\hline
\end{tabular}
\end{center}
\label{table1}
\end{table}%

\begin{figure}[ht]
 \centering
 \includegraphics[width=4.5in]{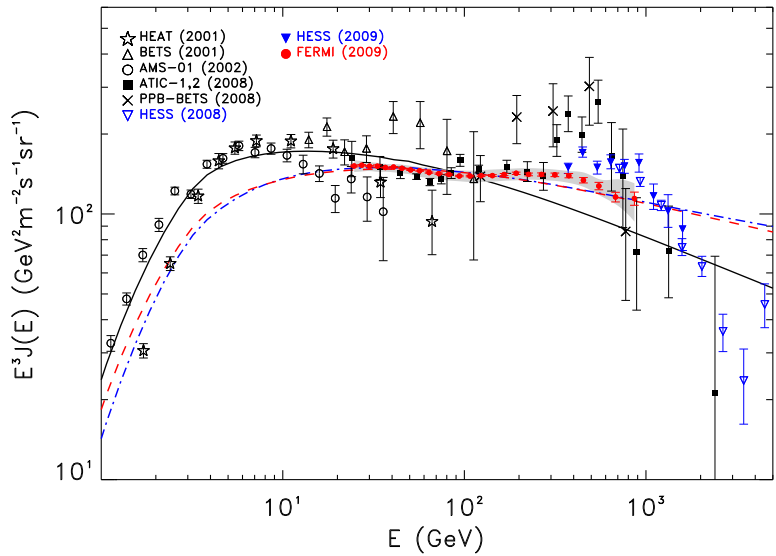}
 \caption{\footnotesize \it In this figure we compare Fermi-LAT CRE data (Abdo et al. 2009 \cite{Fermi_CRE_1}), as well as several other experimental data sets
 (HEAT: Du Vernois et al. 2001 \cite{DuVernois:2001bb};  AMS-01: Aguilar et al. 2002 ~\cite{ams1};  ATIC: Chang et al. 2008 ~\cite{atic}; 
 PPB-BETS:Tori et al. 2008 ~\cite{Torii:2008xu}; H.E.S.S. 2008: Aharonian et al. 2008, ~\cite{hess}; H.E.S.S. 2009 Aharonian  et al. 2009, ~\cite{hess:09}) 
 with the electron plus positron spectrum modeled with GALPROP under the conditions discussed in Sec.\ref{sec:galprop3}.
 The gray band represents systematic errors on the CRE spectrum measured by Fermi-LAT.
 The black continuos line corresponds to the conventional model used in (Strong et al. 2004 \cite{Strong:04}) to fit   
 pre-Fermi data model (model 0 in Tab. 1). 
 The red dashed  (model 1 in Tab. \ref{table1}) and blue dot-dashed lines (model 2 in Tab. \ref{table1}) are
 obtained  with  modified injection indexes in order to fit Fermi-LAT CRE data.
 Both models account for solar modulation using the force field approximation assuming a potential $\Phi = 0.55~{\rm GV}$.     
 }
 \label{fig:elepos_242reac}
\end{figure}

A GALPROP conventional model with $\gamma_0 = 2.54$  was already successfully used to interpret  pre-Fermi CRE data (Strong at al. 2004 ~\cite{Strong:04}).
The other main parameters of that model are reported in the first row of Tab. \ref{table1} (model 0). 
Recently, that model was also shown to reproduce the gamma-ray diffuse emission spectrum measured by Fermi-LAT at intermediate Galactic latitudes (Abdo et al. 2009b\cite{GeVexc}).  The CRE spectrum predicted by that model, however, is  significantly softer than the spectrum measured by Fermi-LAT (see Fig. \ref{fig:elepos_242reac}). 

We find that if $\delta = 1/3$ a conventional model with injection spectral index $\gamma_0 = 2.42$ above 4 GeV (model 1 in Tab. \ref{table1}- red dashed line in Fig. \ref{fig:elepos_242reac}), or if $\delta = 0.6$ and $\gamma_0 = 2.33$  (model 2 in Tab. \ref{table1} - blue dashed line in Fig. \ref{fig:elepos_242reac}),
provide much better fits of Fermi-LAT CRE data. The electron spectrum influences predictions for Galactic diffuse gamma rays via IC and bremsstrahlung emissions.  
This topic will be addressed in a forthcoming paper comparing Fermi-LAT diffuse gamma-ray measurments with models over the whole sky.
Here it suffices to say that the difference between the diffuse gamma-ray spectra predicted with model 0 and model 1 
of Tab. \ref{table1} (based on pre-Fermi data) is not large since the electron injection spectrum change from 2.54 to 2.42 causes a change of only 0.06 in the IC index. Thus the intermediate latitude predictions (Abdo et al. 2009b~\cite{GeVexc}) are hardly affected. 

While, models 1 and 2 provide good representations of the Fermi-LAT data from 20 to 1000 GeV, as shown in Fig.\ref{fig:elepos_242reac} (red and blue dashed lines), they do not fit the AMS-01(Aguilar et al. 2002 ~\cite{ams1}) and HEAT (DuVernois et al. 2001 \cite{DuVernois:2001bb}) data below 20 GeV. Note that our results use a solar modulation potential $\Phi =~550$~MV which is appropriate for the AMS-01 and HEAT data taking periods (Barwick et al. 1997 ~\cite{Barwick:1997ig}).
 This discrepancy may only partially be interpreted in terms of systematic uncertainty on energy calibration, which may result in a +10\%, - 20\% rigid shift of the 
Fermi-LAT data  (see Abdo et al. 2009 ~\cite{Fermi_CRE_1}).  
Some tuning of the theoretical models at low energy may also be required, e.g. by changing the assumptions on solar modulation, or on particle propagation/losses at low energy.  
It should also be noted that all figures showed in this paper have been obtained by assuming $\gamma_0 = 1.6$ below 4 GeV, as done in Strong et al. (2004 \cite{Strong:04})
in order to reproduce the diffuse gamma-ray spectrum measured by CGRO/EGRET and COMPTEL.
Other choices of the source spectral index at those low energies may also be considered which may improve the agreement of the models with low energy pre-Fermi data
without affecting the interpretation of CRE spectrum measured by Fermi-LAT.

The excess in the prediction of the models considered here with respect to H.E.S.S. data above 1 TeV (Aharonian et al. 2008~\cite{hess}) may be a consequence of 
a cutoff in the CRE source spectrum or of the breakdown of the source spatial continuity and steady state hypothesis beyond that energy. This feature is to be expected as a consequence of the rapidly growing IC and synchrotron losses at high energy.  These losses reduce the lifetime of  $\approx 1~\TeV$ electrons to $\sim 5 \times 10^5~{\rm yr}$
implying they can diffuse only a few hundred parsec from their sources. Since such a length is comparable with the mean distance between active SNRs, this may induce significant structure in the high energy part of the electron spectrum  compared to the simple homogeneous models considered in the above. 

Those effects can be accounted for either by following  a `statistical' approach,  which tries to estimate the effect of source stochasticity,  or by trying to model the contribution of actually observed nearby sources. In the following subsection we shortly discuss the former approach leaving a detailed analysis of a particular realization of the latter approach to Sec.\ref{sec:pulsars}.

\subsection{The possible effect of source stochasticity}\label{sec:stochasticity} 

Because of their rapid energy losses at high energy, combined with the stochastic nature of astrophysical sources,  fluctuations may arise
in the locally observed electron spectrum that need to be considered when interpreting the Fermi-LAT electron data.
Those effects can be evaluated either by running GALPROP in 3D mode with stochastic sources (Strong et al. 2001 \cite{Strong:2001qp}) or by means of analytical calculations 
(Pohl \& Esposito 1998~\cite{Pohl:1998ug}). Here we will follow the latter approach (basic equations are given in the Appendix).

The main parameters involved are the frequency of source events as a function of position in the Galaxy and the time over which electrons are injected by each source into the
interstellar medium. Other possible effects are the distribution of spectral shapes over the source population (as traced e.g. by SNR radio spectral indices), and the influence of Galactic spiral structure on the source distribution.

\begin{figure}[ht]
  \centering
  \includegraphics[width=4.5in]{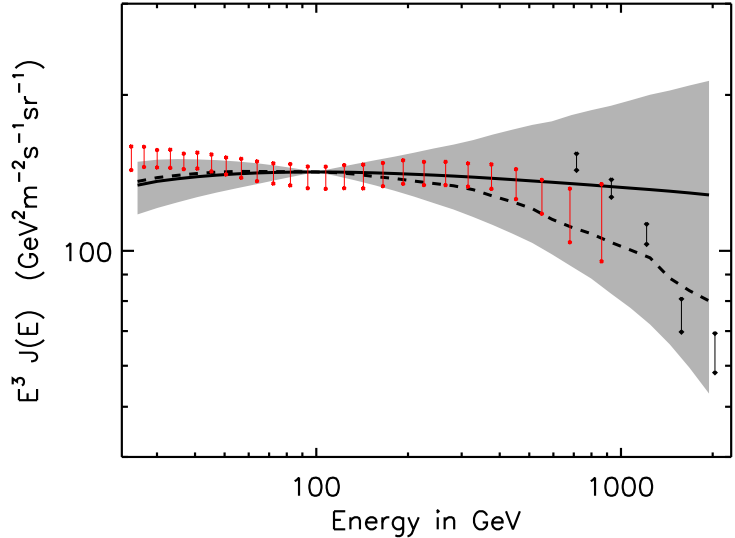}
  \caption{\footnotesize \it 
  Results of an analytical calculation for stochastic sources, including Gould's Belt (see Pohl et al.~2003\  \cite{GouldBelt}).
 The propagation parameters are those of model 1 in Tab. \ref{table1}, and all spectra 
 are normalized to the fiducial flux at 100~GeV. The solid line 
 gives the average spectrum that one would obtain, if the sources were continuously
 distributed. The shaded are indicates the 1-$\sigma$ fluctuation range of the electron 
 flux at each energy. The dashed line indicates one randomly chosen,
 actual electron spectrum.  Fermi-LAT and H.E.S.S. data points are represented in red and black respectively.}
  \label{fig:stochasticity}
\end{figure}

For ease of comparison, we will use the propagation parameters of model 1 in Tab. \ref{table1} and normalize all spectra to the fiducial flux at 100~GeV. 
The main parameters involved are the time period for which electrons are released by each SNR, here 20~kyr, and the rate of supernovae
as a function of location in the Galaxy, for which we use a time-dependent model of supernovae in Gould's Belt superposed on a uniform supernova distribution 
in the Galactic Plane with half-thickness 80~pc (for details see Pohl et al. 2003 et al. \cite{GouldBelt}). Gould's Belt
enhances the local SN rate, resulting in marginally harder electron spectra. Fig.~\ref{fig:stochasticity} shows, for merely illustrative purposes, the result of the analytical calculations. 
We use a soft electron-injection spectrum and a shallow energy dependence of the diffusion coefficient, for which the 
contribution of young and nearby SNR is not efficently truncated at low energy, resulting in
a broad, relatively flat feature in $E^3\,J(E)$. Earlier studies (Pohl \& Esposito 1998 \cite{Pohl:1998ug}) assumed a stronger
energy dependence of the diffusion coefficient, $\delta=0.6$, resulting in a significantly bumpier 
electron spectra.  Shown in Fig.~\ref{fig:stochasticity} is the 1-$\sigma$ fluctuation amplitude in the electron flux.
In 32\% of cases we find the electron flux outside of the shaded band. The corresponding
uncertainty in spectral index can be estimated from the opening angle of the shaded band to be
$\Delta\alpha\simeq 0.2$ between 100~GeV and 1~TeV. 
This implies that the spectrum measured by Fermi could differ by 0.2 from the Galactic average because of such fluctuations.

The solid line gives the average spectrum, which indeed is slightly harder than that shown in Fig.~\ref{fig:elepos_242reac}, solely an effect of Gould's Belt.
The dashed line indicates one randomly chosen, actual electron spectrum which happens to show some curvature so to better match Fermi-LAT and H.E.S.S.
data. 

\subsection{The positron excess problem} 

\begin{figure}[ht]
 \centering
 \includegraphics[width=4.5in]{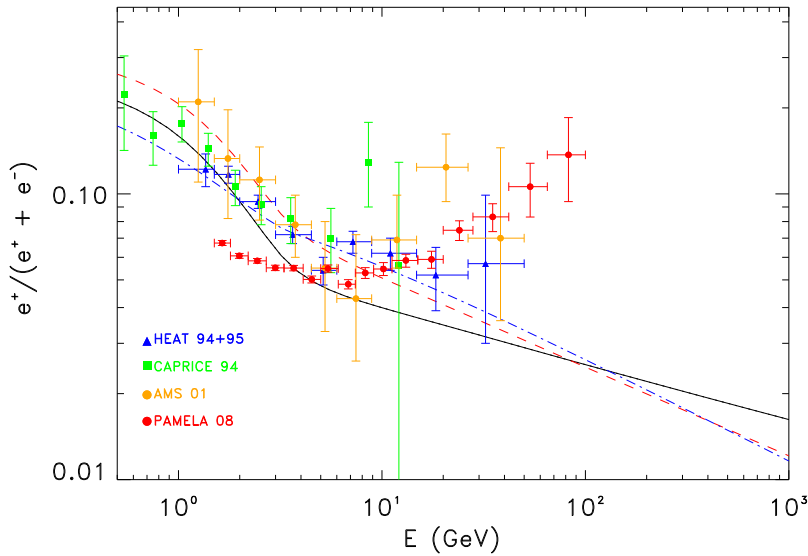}
 \caption{\footnotesize \it 
  In this figure we compare the positron fraction corresponding to the same models used to draw
  Fig. \ref{fig:elepos_242reac} with several experimental data sets (HEAT: Barwick et al. 1997 ~\cite{Barwick:1997ig}; CAPRICE: Boezio et al. 2000 \cite{Boezio:00};
  AMS-01: Aguilar et al. 2002 ~Aguilar et al. 2002 \cite{ams1}; PAMELA: Adriani et al. 2009, 2009b~\cite{PAMELA,PAMELA_Nature}).
  The line styles are coherent with those in that figure. 
  Note that our results use a solar modulation potential  $\Phi = 0.55~{\rm GV}$ which is appropriate for the AMS-01 and HEAT data taking periods 
  (Barwick et al. 1997 \cite{Barwick:1997ig}). 
  It is not appropriate for the PAMELA data taking period, and impacts agreement among the experiments and our model with the PAMELA data below ~ 10 GeV.
 }
  \label{fig:pos_ratio_242reac}
\end{figure}

A serious problem that those GCRE models face is that the positron fraction  $e^+/(e^+ + e^-)$ they predict is not consistent with that measured by PAMELA (Adriani et al. 2009, 2009b~\cite{PAMELA,PAMELA_Nature}). 
 While previous electron data were affected by large uncertainty on the normalization and the slope of the electron background,  to accommodate the PAMELA positron fraction with a steep electron spectrum and standard secondary e$^+$ production was already a hard task (see e.g. Delahaye et al. 2009 ~\cite{Delahaye:2008ua}). 
 Fermi's precise measurement of a hard $\approx E^{-3}$ electron spectrum, further sharpes this discrepancy. 
To modify the standard GCRE models by introducing source stochasticity does not help to predict the PAMELA positron spectrum correctly.  
Reference models 0, 1 and 2 are shown in Fig. \ref{fig:pos_ratio_242reac}. They do not account for the rise in the positron fraction seen by PAMELA, so to explain this data, some additional sources of positrons is required. This situation does not improve by considering other possible combinations of the propagation parameters and of the electron source spectral index that give a good fit to the Fermi-LAT electron spectrum. 

\section{Pulsar Interpretation}\label{sec:pulsars}

Pulsars are undisputed sources of relativistic electrons and positrons, believed to be produced in the magnetosphere and subsequently possibly reaccelerated by the pulsar winds or in the supernova remnant shocks (see e.g.  Shen 1970 \cite{Shen}; 
 Harding \& Ramaty 1987 \cite{Harding:87}; Arons 1996 \cite{Arons:96}, Chi at al. 1996 \cite{Chi:96}; Zhang \& Cheng 2001 \cite{Zhang:01}).  
 For bright young pulsars the maximal acceleration energy can be as large as $10^3$ TeV  (see e.g. Aharonian, 2004 \cite{Aharonian_book}). 
 While this quantity is expected to decrease with the pulsar spin-down luminosity, it may still be well above a TeV for middle-age or, so called, {\it mature}
 pulsars (i.e. with age $10^4 \simleq T \simleq 10^6~{\rm yr}$ )  (see e.g.  B\"ushing  2008,2008b  \cite{Busching:2008zzb,Buesching:2008hr} and ref.s therein). 
 As noted in Chi at al. (1996 ~\cite{Chi:96}) electron and positron pairs accelerated in the PWNe  should be confined in the nebula itself or the surrounding 
supernova remnant and  accumulate there until those systems merge into the ISM,  $10^{4}$ - $10^{5}$ years after the pulsar birth
(for a review on PWN see e.g. Gaensler 2006 \cite{Gaensler:2006ua}).   Since this process is expected to be relatively fast, and the following pulsar emission to be
unimportant (as the spin-down power decreases like ${\dot E}_{\rm PSD} \propto t^{-2}$ approximatively)  {\it mature} pulsars can effectively be treated as burst-like sources 
of electrons and positrons.  

At energies between $100~\GeV$ and $1~\TeV$ the electron flux reaching the Earth may therefore be the sum of an almost homogeneous and isotropic GCRE 
component produced by Galactic supernova remnants and the local contribution (LCRE) of  a few pulsars (SNRs), with the latter expected to contribute more and more significantly as the energy increases.   In order to account for this possibility, 
here we sum the analytically computed electron and positron spectrum from observed pulsars in the nearby  to the GCRE component computed with GALPROP. 
Our approach is similar to that followed in Aharonian et al. (1995 \cite{Aharonian:95}),  Atoyan et al. (1995 \cite{Atoyan:95}) and Kobayashi et al. (2004 \cite{Kobayashi:2003kp}).

We do not consider the discrete contribution of nearby conventional shell-type SNR to the high energy CRE flux (which is understood to be included in the GCRE component). 
The main reason for such a choice is that the interpretation of Fermi data discussed in the following requires
an electron (and  positron) injection spectrum which is significantly harder than generally expected for conventional shell-type SNRs (see, however, Blasi 2009 \cite{Blasi:2009hv} for a different interpretation of Fermi and PAMELA results based on secondary CR acceleration in the SNRs). 

We compute the spectrum of electrons and positrons from each pulsar by following the approach reported in the Appendix.
The basic input is the  $e^\pm$ energy release of each mature pulsar that we determine by integrating the observed spin-down luminosity over time giving 
 (see e.g. Profumo 2008 ~\cite{Profumo:2008ms})
\begin{equation}
\label{fig:pulsar_energy}
E_{e^\pm}  \simeq \eta_{e^\pm}~  {\dot  E}_{\rm PSD}~\frac{T^2}{\tau_0}~,
\end{equation}
where $ {\dot  E}_{\rm PSD}$ is the present time spin-down luminosity determined from the observed pulsar timing, $T = P/2{\dot P}$ (where $P$  is the pulsar period) 
the pulsar age,  and $\eta_{e^\pm}$ is the $e^\pm$ pair conversion efficiency of the radiated electro-magnetic energy. For the characteristic luminosity decay time we assume 
$\tau_0 = 10^4~{\rm years}$ as conventionally adopted for mature pulsars (see e.g. Aharonian et al.1995 \cite{Aharonian:95}). 

A relevant parameter determining the shape of the high energy part of the electron spectrum is the ratio between the injection cutoff
to the maximal arrival energy allowed by energy losses during propagation to the Earth  $\epsilon \equiv  E_{\rm cut}/E_{\rm max}$. If
$\epsilon > 1$  the exponential cutoff at the source plays no role and the electron spectrum due to a single nearby source should be
sharply suppressed above $E_{\max}$ (see last term in Eq. (\ref{eq:pulsar_spectrum})).  In the opposite case
($\epsilon \ll 1$) a significantly smoother cutoff is expected. This may play a relevant role when trying to disentangle pulsar from dark matter
interpretations of recent CRE data (see Sec. \ref{sec:discussion}). 

The setup we use here to model the large-scale GCRE spectrum adopts $\gamma_0 = 2.54$ and $\delta = 0.33$.
This model is very similar to the conventional GALPROP model (Strong et al. 2004~\cite{Strong:04}) (model 0 in Tab. \ref{table1})  which has been successfully used to model the diffuse gamma-ray spectrum measured by Fermi at intermediate Galactic latitudes (Abdo et al. 2009b \cite{GeVexc}).  With respect to that reference model, however,  the primary electron spectrum normalization needs to be slightly reduced (by a factor $\sim 0.95$) to leave room to the additional pulsar component. 
We verified  that such tuning has a small effect on the diffuse gamma-ray spectrum. This is not unexpected since, for conventional models, above $0.1~\GeV$  the IC and bremsstrahlung contributions at intermediate latitudes are less important than the hadronic and the extra-Galactic components.  
The other relevant propagation parameters are set to match the nuclear CR data; different choices of the value of $\delta$ (e.g. $\delta = 0.6$)  would not affect significantly our results.

\subsection{The contribution of nearby pulsars} \label{sec:pulsar}

For illustrative purposes, in this subsection we consider the case in which only few nearby pulsars, and only for a representative choice of the relevant parameters,  
contribute significantly to the high energy electron flux reaching the Earth. A more realistic analysis accounting for the contribution of more distant sources, 
and parameter variance, will be performed in Sec.\ref{sec:multiple_pulsars}.

We select candidate sources from the ATNF radio pulsar catalogue \footnote{http://www.atnf.csiro.au/research/pulsar/psrcat/}  (Manchester et al., 2005 \cite{Manchester:05}). 
We require a distance smaller  than $d < 1~\kpc$ and an age larger than $T > 5 \times 10^4~{\rm years}$.  As explained at the beginning of Sec.\ref{sec:pulsars} younger pulsars, 
like Vela ($d = 290~\pc$, $T =  1.1 \times 10^4~{\rm years}$),  are not expected to play any role here since their electrons should be still confined in the PWN or in the SNR envelope. 

Among this set of candidate sources we found that only the Monogem  (PSR B0656+14) at a distance of $d = 290~\pc$  and age
$T =  1.1\times 10^5~{\rm years}$ (Manchester et al. 2005), and the Geminga pulsar (PSR J0633+1746) with $d = 160~\pc$  and $T =   3.7\times 10^5~{\rm years}$ 
(Caraveo et al. 1996 ~\cite{Caraveo:96}) give a significant contribution to the high energy electron and positron flux reaching the Earth. 
The observed spin-down luminosities of these pulsars are ${\dot  E}_{\rm PSD} \simeq 3.8 \times 10^{34}~\erg~ \s^{-1}$ and ${\dot  E}_{\rm PSD} \simeq 3.2 \times 10^{34}~\erg~ \s^{-1}$ respectively. 

\begin{figure}[ht]
 \centering
 \includegraphics[width=4.5in]{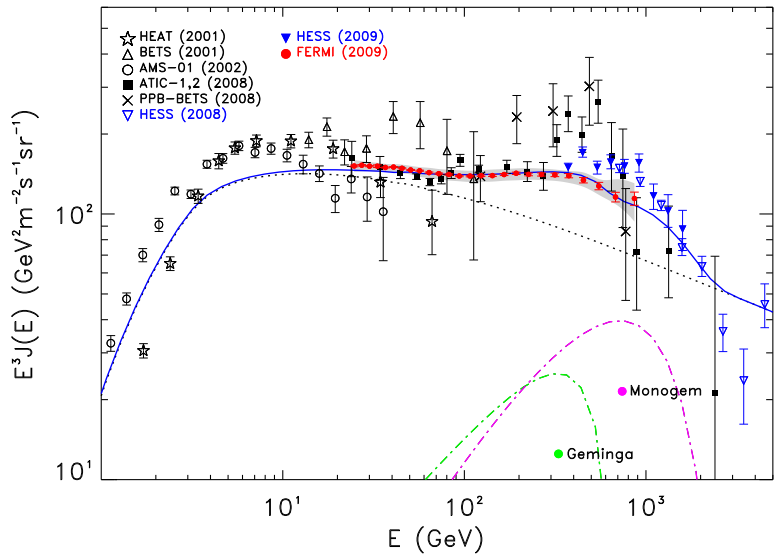}
 \caption{\footnotesize \it In this figure we represent the electron-plus-positron spectrum (blue continuos line) computed 
 in a case in which only observed pulsars from the ATNF catalogue (Manchester 2005 \cite{Manchester:05}) with distance $d < 1~\kpc$ 
 plus the large-scale Galactic component  (GCRE) give a significant contribution. 
 The dominant contribution of Monogem and Geminga pulsars, analytically computed for a representative choice of the relevant parameters (see text)
 is shown as colored dot-dashed lines, while the GCRE,  computed with GALPROP is shown as a black-dotted line.
 The gray band represents systematic errors on the CRE Fermi-LATdata. 
 Solar modulation is accounted as done in Fig.\ref{fig:elepos_242reac}. 
 }
 \label{fig:elepos_Monogem_242reac}
\end{figure}

\begin{figure}[ht]
 \centering
 \includegraphics[width=4.5in]{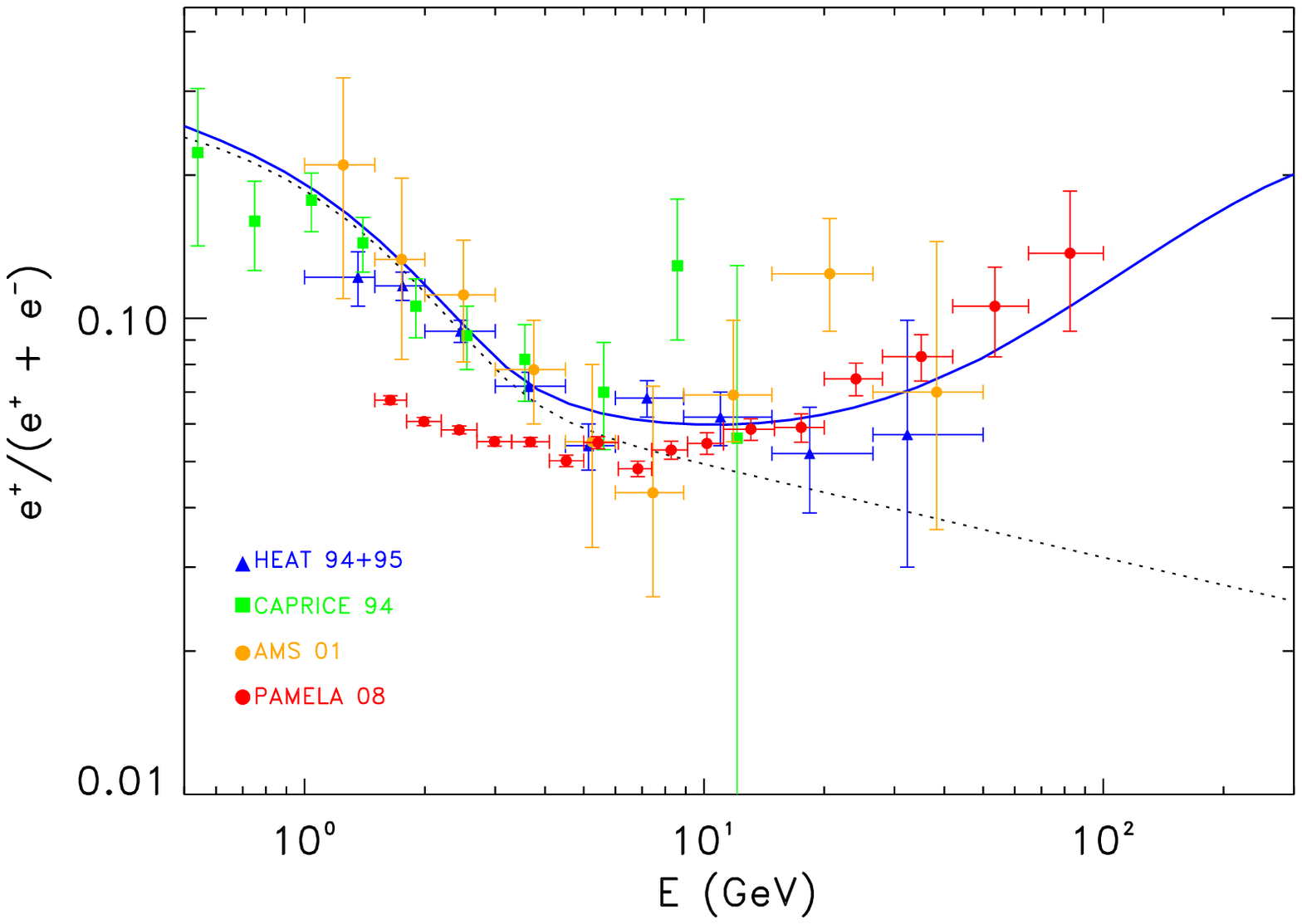}
 \caption{\footnotesize \it The positron fraction for the same scenario as in Fig.\ref{fig:elepos_Monogem_242reac}. 
 Line styles are coherent with those in that figure.
 Solar modulation is accounted as done in Fig.\ref{fig:pos_ratio_242reac}.} 
  \label{fig:pos_ratio_Monogem_242reac}
\end{figure}

In (Hooper at al. 2008, \cite{Hooper:2008kg}) the authors showed that PAMELA (Adriani et al. 2009, 2009b~\cite{PAMELA,PAMELA_Nature}) positron fraction data can be fitted under the 
hypothesis that either Monogem or Geminga inject electron-positron pairs with a spectrum of the form given in Eq.(\ref{eq:inj_spectrum}) with $\Gamma =1.5$
\footnote{Note that in (Hooper at al. 2008, \cite{Hooper:2008kg}) the authors used a simplified version of Eq. (\ref{eq:inj_spectrum}). While this plays no role interpreting PAMELA data, as done in that paper, using the exact expression given in Eq.(\ref{eq:inj_spectrum}) is necessary here in order to correctly model the electron spectrum above few hundred GeV.}.  
Here we find a similar result.  The plots in Fig.s \ref{fig:elepos_Monogem_242reac} and \ref{fig:pos_ratio_Monogem_242reac} are drawn using the source spectral index $\Gamma = 1.7$. 
We notice here that such a value is compatible with the synchrotron emission spectra observed by pulsar radio observations as well as with 
gamma-ray spectra measured by EGRET in the $0.1-10$ GeV range (Thomson et al. \cite{Thompson:1994sg}) which loosely constrains it in the range 
$1.4 < \Gamma < 2.2$.  In particular, in the case of  Crab PWN,  it was shown that gamma-ray measurements can be interpreted in terms of IC emission 
from a population of electrons having a power-law spectrum with $\Gamma \simeq 1.5$ up to $\sim 200~\GeV$, becoming steeper at higher energies, which is very
close to that value used here. Since the PWN magnetic field, hence synchrotron energy losses, decrease with the pulsar age, that break is expected to be at much larger 
energies for mature pulsars (see e.g. Aharonian et al. 1997 \cite{Aharonian:97}).   

The cutoff energy $E_{\rm cut}$ for mature pulsars is unknown. For young pulsars the PWN gamma-ray spectra observed by ACTs have been interpreted in terms of IC 
emission of electrons with  $E_{\rm cut} \approx 10^3~\TeV$ (see e.g. Aharonian 2004~\cite{Aharonian_book}).  That quantity, however, is expected to be considerably smaller for older pulsars 
as it decreases with the pulsar spin-down luminosity (see e.g. B\"ushing et al.  2008 \cite{Busching:2008zzb,Buesching:2008hr}).  

It is evident from Fig.s \ref{fig:elepos_Monogem_242reac} and \ref{fig:pos_ratio_Monogem_242reac}  that PAMELA and Fermi-LAT CRE data can nicely be fit under
the same conditions. 
Lacking a fully consistent theory allowing the prediction of the cutoff energy $E_{\rm cut}$ and the efficiency $\eta_{e^\pm}$ as a function of pulsar age and luminosity, we 
assume  here that both pulsars share the same values of those parameters and tune them to fit the data (this choice is, however,  not critical in order to reproduce the data). 
For both pulsars we also assume the same delay $\Delta t$  between the pulsar birth and the electron delivery in the ISM.  

We find that our predictions are in remarkable agreement with the whole set of data \footnote{The small deficit respect to H.E.S.S. data is within the systematic errors
reported for that experiment.} for several combinations of  those parameters. In particular  
the diagrams shown in Fig. \ref{fig:elepos_Monogem_242reac} and \ref{fig:pos_ratio_Monogem_242reac}  have been obtained with  
$E_{\rm cut} = 1100~\GeV$,  $\eta_{e^\pm} = 40 \% $  and $\Delta t = 6 \times 10^4~{\rm yr}$.  As discussed above, these values are compatible with our 
 knowledge of particle acceleration in PWNe.  It is understood that our choice of the parameters represents a particular realization of the scenario discussed in this
 section. The effect of changing those parameters in a reasonable range is discussed below.  

\subsection{Including the contribution of distant pulsars} \label{sec:multiple_pulsars}

In a more realistic scenario the flux of high energy electrons and positrons reaching the Earth will receive a contribution from a large number of pulsars. 
Modeling the electron spectrum is this case is more difficult due to parameter variance among the different pulsars and the possible contribution of unobserved pulsars. 
Nevertheless, it is important to verify how the successful illustrative interpretation discussed in the previous section may change under more reasonable assumptions.   

For this purpose we sum the contribution to the electron and positron flux of all known mature pulsars in the ATNF radio pulsar catalogue
within a distance larger than that considered in the previous subsection, namely $d < 3~\kpc$  and age $T > 5 \times 10^4~{\rm yr}$ ( $\sim 150$  pulsars).  
This set includes Monogem and Geminga. The contribution from more distance pulsars is negligible in the energy range considered.
For these pulsars we use the spin-down luminosities given in the catalogue and randomly vary the relevant parameter in the following ranges:
$800 < E_{\rm cut} < 1400~\GeV$,  $10 < \eta_{e^\pm} <  30~\%$  and $5 <  (\Delta t / 10^4~{\rm yr} ) <  10$ and $1.5 < \Gamma < 1.9$.
Since, respect to the previous subsection, a larger number of pulsars contribute to the observed electron spectrum, on average,  smaller values of the 
 $e^\pm$ conversion efficiency are required here.  In all cases we use the same model for the GCRE component (model 0 in Tab. \ref{table1} rescaled by 0.95) as discussed 
at the beginning of  Sec.\ref{sec:pulsars}.

\begin{figure}[ht]
  \centering
  \includegraphics[width=4.5in]{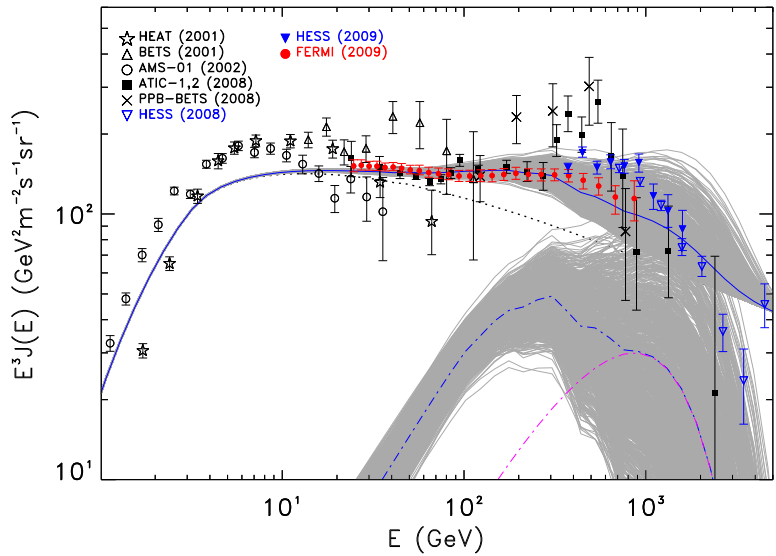} 
  \caption{\footnotesize \it In this figure we compare the electron plus positron spectrum 
  from multiple pulsars plus the Galactic (GCRE) component with experimental data (dotted line). We consider 
  the contribution of all nearby pulsars in the ATNF catalogue with $d < 3~ {\rm kpc}$ 
  with age  $5\times 10^4 < T \,< \, 10^7 \, {\rm yr}$ by randomly varying  $E_{\rm cut}$, $\eta_{e^\pm}$ 
   $\Delta t$ and $\Gamma$ in the range  of parameters given in the text. Each gray line represents the sum of all 
   pulsars for a particular combination of those parameters.  The blue dot-dashed (pulsars only) and blue solid lines (pulsars + GCRE component)
   correspond to a representative choice among that set of possible realizations.  
  The purple dot-dashed line represents the contribution of Monogem pulsar in that particular case.
  Note that for graphical reasons here Fermi-LAT statistical and systematic errors are added in quadrature. 
  Solar modulation is accounted as done in previous figures.
  }
  \label{fig:elepos_random_pulsars}
\end{figure}

\begin{figure}[ht]
  \centering
  \includegraphics[width=4.5in]{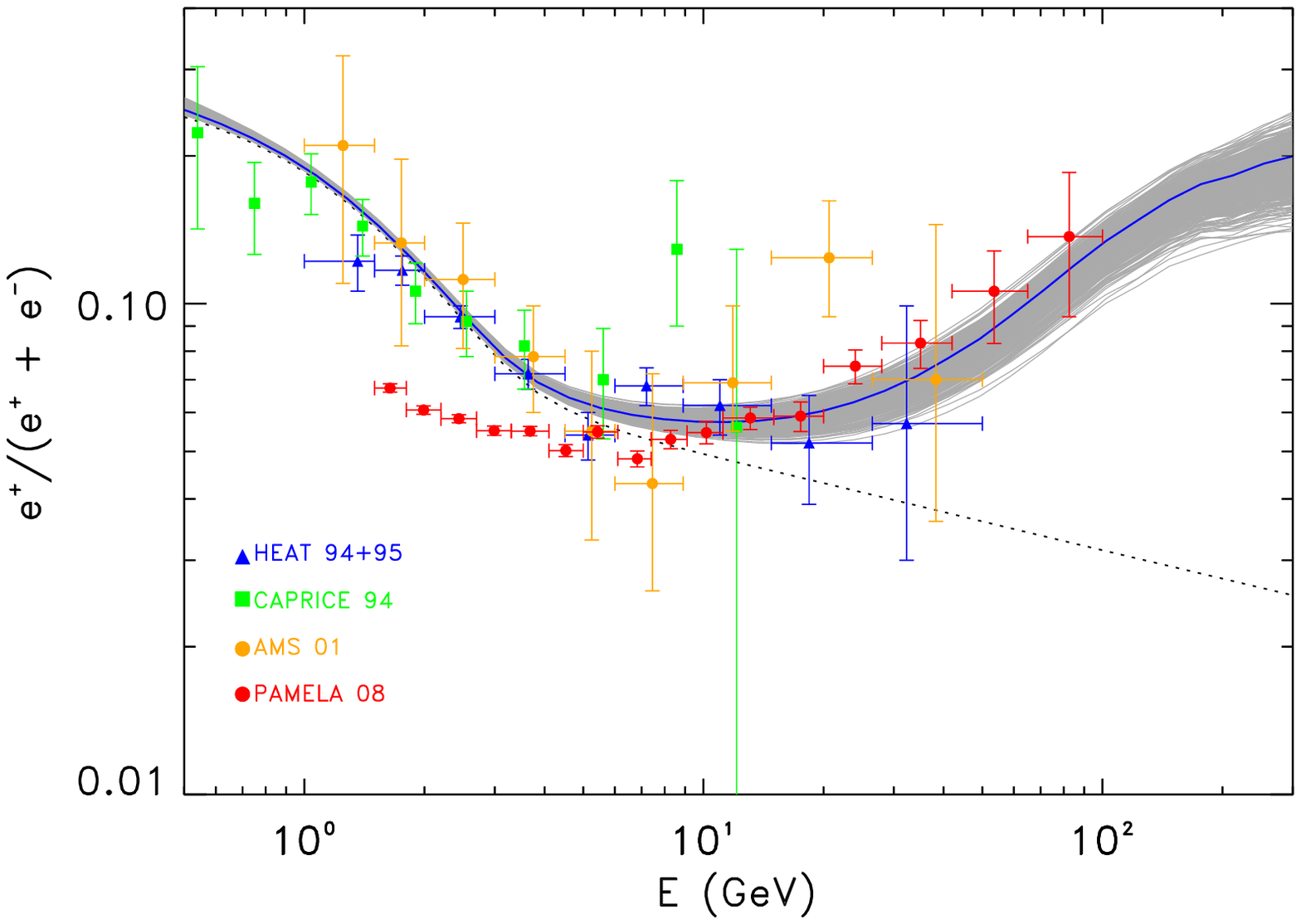}
  \caption{\footnotesize \it  The positron fraction corresponding to the same models used to draw 
   Fig. \ref{fig:elepos_random_pulsars} is compared with several experimental data sets.
   The line styles are coherent with those in that figure. 
   Solar modulation is are accounted as done in  }
  \label{fig:pos_ratio_random_pulsars}
\end{figure}

We find that  Fermi-LAT CRE data, as well as PAMELA positron ratio data, comfortably lay within the bands of those realizations (see Fig.~\ref{fig:elepos_random_pulsars}) and are compatible with the positron fraction measured by PAMELA (see Fig. \ref{fig:pos_ratio_random_pulsars}) .  
 
It should be noted that the ATFN catalogue does not include all pulsars. Some pulsars radio beams are not pointing toward us and also selection effects in the radio detection intervene to reduce the number of the observed  pulsars. Furthermore, the recent discovery of a population of radio-quiet gamma-ray pulsars by Fermi-LAT (Abdo et al. 2009c \cite{blind_search}) has demonstrated that those pulsars are a significant fraction of the total pulsar set. 
 We do not expect, however, that the average CRE spectral shape would change significantly by accounting for pulsars not included in the ATFN catalogue.
The larger electron and positron primary flux due to the contribution of those sources can be compensated by invoking a smaller pair conversion efficiency 
$\eta_{e^\pm}$ making this scenario even more appealing. 
While selection effects may lead to underestimate older pulsar at large distance, their role is almost negligible shaping the CRE high energy spectrum. 
 The role of unobserved pulsars is, however, more important in the PAMELA energy range. In order to account for those objects, an alternative approach, is to use an average 
pulsar contribution from the Galactic disk rather than the catalogue (see e.g. Hooper at al. 2008, \cite{Hooper:2008kg}). Indeed by following a similar approach we found that adding 
the contribution of ATNF pulsars for $r < 1~\kpc$ to an average contribution from more distant pulsars, again both Fermi-LAT and PAMELA results can be consistently be reproduced with a $\sim 10\%$  electron and positron conversion efficiency. 

In principle, fluctuations in the Galactic CRE component should also be considered (as discussed in Sec. \ref{sec:GCRE}). This, however, may only increase
 the parameter ranges that are compatible with data.  The apparent discrepancy between our prediction and the H.E.S.S. data above  2 TeV may indeed be explained
as a consequence of those fluctuations. Therefore we conclude that under reasonable assumptions, the Fermi-LAT data on the electron spectrum, the H.E.S.S. data (within their systematic errors), and positron fraction PAMELA data are all consistent with the pulsar emission of electrons and positrons scenarios discussed here.

\section{Dark matter interpretation}\label{sec:DM}

We discuss here the possibility that the measured CRE data (including the PAMELA positron fraction measurement) originate from the pair-annihilation or from the decay of Galactic dark matter. We assess the impact of the Fermi-LAT data on the flux of energetic $e^\pm$ for what shall be referred to as the ``dark matter interpretation'' of the reported spectra. We focus here on the scenario of dark matter pair-annihilation. 

The new Fermi-LAT data affect a dark matter interpretation of CRE data in at least three ways:
\begin{enumerate} 
\item The rationale to postulate a particle dark matter mass in the 0.5 to 1 TeV range, previously motivated by the ATIC data and the detected ``bump'', is now much weaker, if at all existent, with the high statistics Fermi-LAT data; 
\item  CRE data can be used, in the context of particle dark matter model building, to set constraints on the pair annihilation rate or on the decay rate, for a given dark matter mass, diffusion setup and Galactic halo model;
\item as discussed in Sec.\ref{sec:GCRE}, unlike the Fermi-LAT CRE result, the PAMELA positron fraction measurement requires one or more additional primary sources in addition to the standard GCRE component, as discussed in Sec.~\ref{sec:GCRE}; if the PAMELA data are interpreted in the context of a dark-matter related scenario, Fermi-LAT data provide a correlated constraint to the resulting total CRE flux.
\end{enumerate}

We emphasize here that, although not {\em per se} needed from data, a dark matter interpretation of the Fermi-LAT and of the PAMELA data is an open possibility.  
Nevertheless we note that a dark matter interpretation of the Fermi-LAT data is disfavored for at least the three following reasons:
\begin{itemize}
\item Astrophysical sources (including pulsars and supernova remnants) can account for the observed spectral features, as well as for the positron ratio measurements (sec.~\ref{sec:pulsar}): no additional exotic source is thus required to fit the data, although the normalization of the fluxes from such astrophysical objects remains a matter of discussion, as emphasized above.
\item Generically, dark matter annihilation produces antiprotons and protons in addition to $e^\pm$. If the bulk of the observed excess high-energy $e^\pm$ originates from dark matter annihilation,  the antiproton-to-proton ratio measured by PAMELA (Adriani et al. 2009 \cite{Adriani:2008zq}) sets very stringent constraints on the dominant dark matter annihilation modes, as first pointed out by Donato et al. 2009 \cite{Donato:2008jk} (see also Cirelli et al. 2009~ \cite{Cirelli:2008pk}). In particular, for ordinary particle dark matter models, such as neutralino dark matter (Jungman 1996 
\cite{Jungman:1995df}) or the lightest Kaluza-Klein particle of Universal Extra-Dimensions (Hooper \& Profumo 2007 \cite{Hooper:2007qk}), the antiproton bound rules out most of the parameter space where one could explain the anomalous high-energy CRE data.
\item Assuming particle dark matter is weakly interacting, and that it was produced in the early Universe via an ordinary freeze-out process involving the same annihilation processes that dark matter would undergo in today's cold universe, the annihilation rate in the Galaxy would be roughly two orders of magnitude too small to explain the anomalous $e^\pm$ with dark matter annihilation; while this mismatch makes the dark matter origin somewhat less appealing, relaxing one or more of the assumptions on dark matter production and/or on the pair annihilation processes in the early Universe versus today can explain the larger needed annihilation rate; similarly, a highly clumpy Galactic dark matter density profile, or the presence of a nearby concentrated clump, can also provide sufficient enhancements to the rate of dark matter annihilation 
\end{itemize}
Notwithstanding the above caveats, the focus of the present study is to assess the impact of the new Fermi-LAT data on a dark matter interpretation of the excess high-energy $e^\pm$.

We assume for the dark matter density profile $\rho_{\rm DM}$ an analytic and spherically-symmetric interpolation to the results of the high-resolution Via Lactea II N-body simulation (Diemand et al. 2008 \cite{Diemand:2008in}), namely:
\begin{equation}\label{eq:vl2}
\rho_{\rm DM}(r) = \rho_\odot \left(\frac{r}{R_\odot}\right)^{-1.24}\left(\frac{R_\odot + R_s}{r + R_s}\right)^{1.76}\;\;,
\end{equation}
where $\rho_\odot = 0.37\;  \GeV \cdot \cm^{-3}$ is the local density, $R_\odot = 8.5 \; \kpc$ is the distance between the Sun and the Galactic center and $R_s = 28.1\;\kpc$ is a scale parameter. For simplicity, we neglect the effect of clumpiness, as well as the possibility of a nearby, dense dark matter sub-halo. We warn the reader, though, that, while unlikely (Bringmann et al. 2009 \cite{Bringmann:2009ip}), the latter possibility is of great relevance not only for the normalization of the $e^\pm$ produced by dark matter annihilation, but also for the spectral shape (Brun et al. 2008 \cite{Brun:2009aj}).

For illustrative purposes, we focus the present analysis on three simple benchmark classes of models where the flux of antiprotons are generically suppressed to a level compatible with the PAMELA antiproton data (Adriani et al. 2009 \cite{Adriani:2008zq}). Specifically we consider:

\begin{enumerate}
\item {\em Pure $e^\pm$ models}: we define this class of models as one where dark matter annihilation yields a pair of monochromatic $e^\pm$, with injection energies equal to the mass of the annihilating dark matter particle. Notice that dark matter models where the annihilation proceeds into pairs of light intermediate scalar, pseudo-scalar or vector particles $\phi$, subsequently decaying into light fermion (and possibly only $e^\pm$) pairs (see e.g. Finkbeiner and Weiner 2007 \cite{Finkbeiner:2007kk}, Pospelov et al. 2008 \cite{Pospelov:2007mp} and Arkani-Hamed et al. 2008 \cite{ArkaniHamed:2008qn}), produce a different spectrum from the monochromatic $e^\pm$ injection we consider here. Specifically, the resulting $e^\pm$ injection spectra have a further dependence on the mass of the intermediate particle $\phi$. For simplicity, and in order to maintain our discussion at a phenomenological and model-independent level, we do not consider this possibility here. 
\item {\em Lepto-philic models}: here we assume a democratic dark matter pair-annihilation branching ratio into each charged lepton species: 1/3 into $e^\pm$, 1/3 into $\mu^\pm$ and 1/3 into $\tau^\pm$. In this class of models too antiprotons are not produced in dark matter pair annihilation. Examples of models where the leptonic channels largely dominate include frameworks where either a discrete symmetry or the new physics mass spectrum suppresses other annihilation channels (Fox \& Poppitz 2008 \cite{Fox:2008kb}, 
Harnik \& Kribs 2008 \cite{Harnik:2008uu})
\item {\em Super-heavy dark matter models}: As pointed out in Cirelli et al. (2009 \cite{Cirelli:2008pk}), antiprotons can be suppressed below the PAMELA measured flux if the dark matter particle is heavy (i.e. in the multi-TeV mass range), and pair annihilates e.g. in weak interaction gauge bosons. Models with super-heavy dark matter can have the right thermal relic abundance, e.g. in the context of the minimal supersymmetric extension of the Standard Model, as shown in Profumo (2005 \cite{Profumo:2005xd}).
\end{enumerate}

Notice that other dark matter models (including e.g. TeV-scale dark matter particles annihilating in muon-antimuon final states, either monochromatically or through the decays of intermediate particles $\phi$) offer additional possible case-studies, as discussed in Bergstrom et al. (2009 \cite{Bergstrominprep}). We employ here the diffusion model outlined in Baltz \& Edsjo (1999 \cite{Baltz:1998xv}), which is a refined semi-analytic model based on Green functions solutions to the
standard diffusion-loss equation for the space density of charged cosmic rays with free escape boundary conditions and cylindrical symmetry (see
 Baltz \& Edsjo 1999 \cite{Baltz:1998xv} for details). This treatment is fully consistent with the GALPROP setup employed above, but specializes to the case of
emission from Dark Matter annihilation. The diffusive region, the diffusion coefficient and its behavior with particle rigidity and the large scale Galactic CR electron and positron spectrum are the same as for model 0 in Tab. \ref{table1}, rescaled by an overall 0.95,  as also adopted in Sec.\ref{sec:pulsars}.
 We also cross-checked that our results are not qualitatively affected if we used the three MIN, MED and MAX diffusion models of Delahaye et al. (2008 ~\cite{Delahaye:2008ua}), where the label indicates a larger or smaller value of the diffusion coefficient, resulting in a larger or smaller effect on the propagation of the high-energy $e^\pm$. For those three models, the height of the diffusive halo and the dependence of the diffusion coefficient with energy change significantly, while still being compatible with primary-to-secondary ratio measurements.

\begin{figure}[!t]
 \centering
  \includegraphics[width=4.5in]{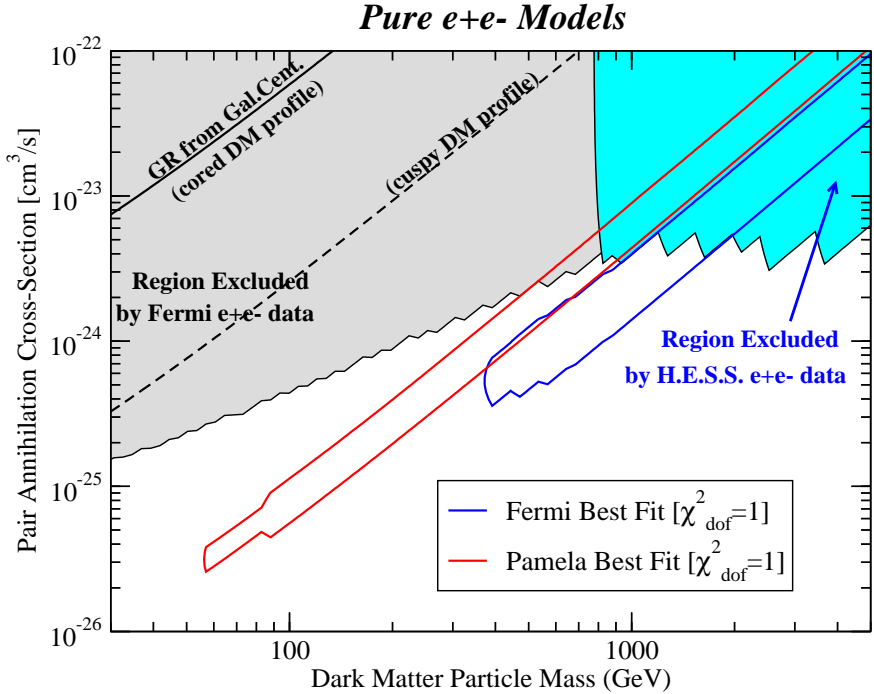}
 \caption{\footnotesize  \it The parameter space of particle dark matter mass versus pair-annihilation rate, for models where dark matter annihilates into monochromatic $e^\pm$.    
 Models inside the regions shaded in gray and cyan over-produce $e^\pm$ from dark matter annihilation with respect to the Fermi-LAT and H.E.S.S. measurements, at 
 the 2-$\sigma$ level. The red and blue contours outline the regions where the $\chi^2$ per degree of freedom for fits to the PAMELA and Fermi-LAT data is at or below 
 1. The solid and dashed line indicate a gamma-ray emission from the Galactic center which should be detectable with Fermi-LAT, assuming, respectively, a cored and
  a cuspy inner slope for the dark matter density profile.}
\label{fig:mod1}
\end{figure}
\begin{figure}[!t]
 \centering
 \includegraphics[width=4.5in]{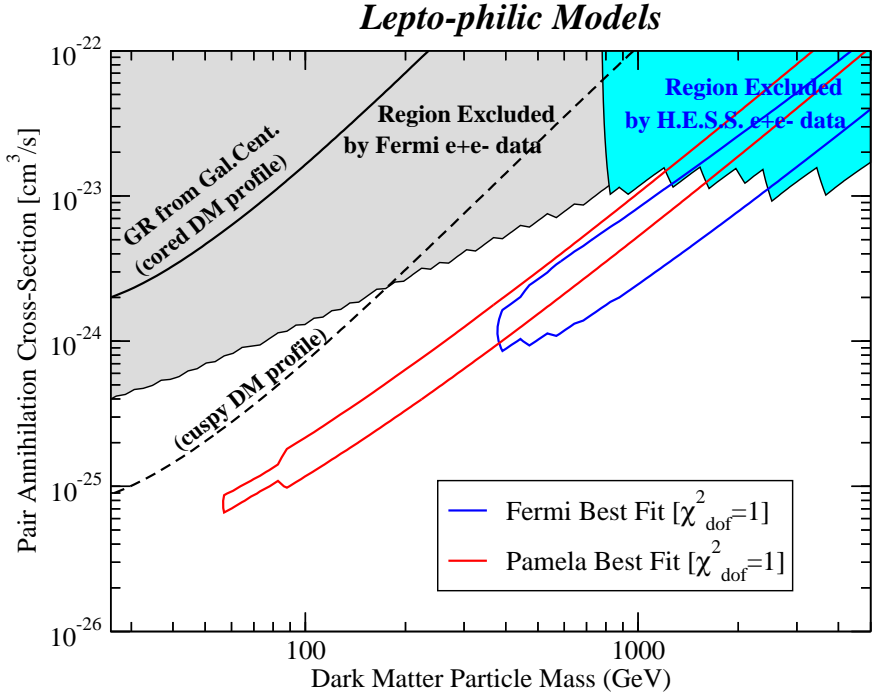}
 \caption{\footnotesize \it As in Fig.~\ref{fig:mod1}, but for ``lepto-philic'' models, where the dark matter pair annihilates democratically into the three charged lepton species.}
 \label{fig:mod2}
\end{figure}
\begin{figure}[!t]
 \includegraphics[width=4.5in]{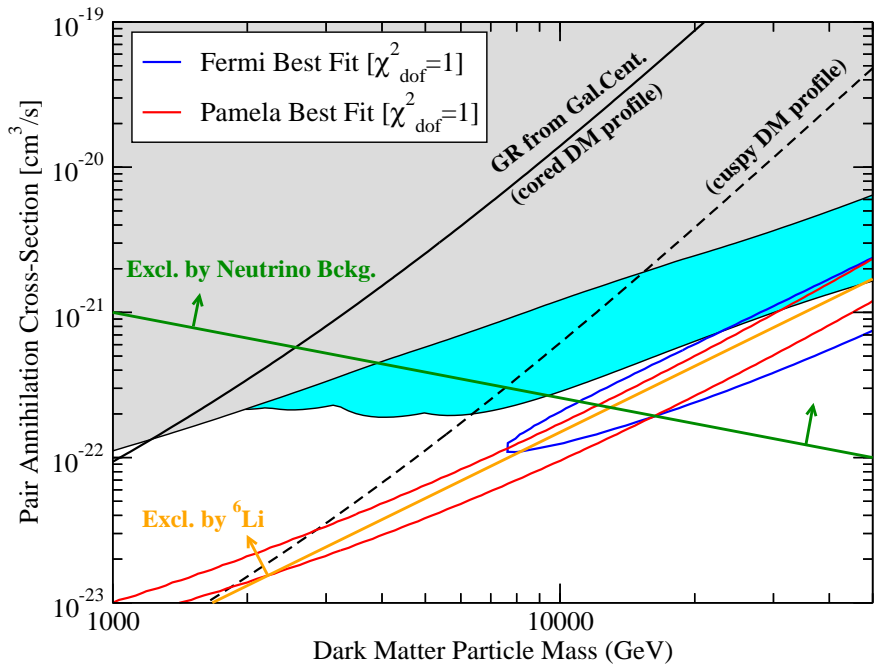}
 \centering
 \caption{\footnotesize \it As in Fig.~\ref{fig:mod1}, but for models featuring heavy dark matter pair-annihilating into electro-weak gauge bosons. The region above the orange line
  is in conflict with the observed primordial ${}^6$Li and ${}^7$Li abundances, and that above the green line is in tension with measurements
   of the neutrino background. As for the previous two figures, the cyan region is excluded by H.E.S.S. $e^\pm$ data \cite{hess}.}
 \label{fig:mod3}
\end{figure}

We show our results for pure $e^\pm$ models in Fig.~\ref{fig:mod1}, for ``lepto-philic'' models in Fig.~\ref{fig:mod2}, and for models with a super-heavy dark matter particle annihilating in electro-weak gauge bosons in Fig.~\ref{fig:mod3}. In each figure, we conservatively shade regions of parameter space inconsistent with the Fermi-LAT CRE data (gray-shading) \cite{Fermi_CRE_1} and/or with the data on the $e^\pm$ flux reported by the H.E.S.S. telescope (Aharonian 2009 \cite{hess}) (cyan-shading). By {\em inconsistent} we mean that even neglecting any other source of $e^\pm$, the flux resulting from dark matter annihilation alone results in an excess of more than 2-$\sigma$ above the measured values.  Notice that the indentations in the edges of the grey and cyan regions only depend on not having considered the smearing in the measured spectrum due to instrumental finite energy resolution. Including this effect does not change, for the Fermi region, the location of the favored and excluded regions. For each value of the dark matter mass and pair annihilation rate, we then calculate the $\chi^2$ to the PAMELA\footnote{In calculating the $\chi^2$ for PAMELA we consider the six higher energy PAMELA bins, with energies larger than $\sim$10 GeV. and Fermi-LAT 
data. Points inside the red and blue regions have a $\chi^2$ per degree of freedom $\simleq 1$  for the PAMELA and Fermi-LAT data sets. }

We also assess the flux of gamma-rays resulting from dark matter annihilation at the center of the Galaxy. While the discrimination of any signature of dark matter annihilation with gamma-ray data relies on the detailed understanding of astrophysical ``backgrounds'', as well on the dark matter density profile, we assume here that a bright enough source at the center of the Galaxy can be discriminated from an astrophysical counterpart even with a spectral analysis of the first year Fermi data (Jeltema 2008 ~\cite{Jeltema:2008hf}). Conservatively, we indicate the lines on the dark matter parameter space where the flux of gamma-rays from the center of the Galaxy corresponds to $10^9$ photons per ${\rm cm}^2
\s^{-1}$.  While in the case of electron-positron and muon pair final states the gamma rays resulting in a dark matter annihilation event only originate from final state radiation (internal bremsstrahlung) off of the light charged leptons, both in the case of $\tau^+ \tau^-$ and of $W^+W^-$ a sizable contribution stems from hadronic channel, namely from the $\pi^0\to\gamma\gamma$ decays of neutral pions produced in the hadronization of strongly interacting decay products of $\tau$'s and $W$'s. While we do not account for the constraints from secondary radiation from subsequent energy losses of electrons and positrons from dark matter annihilation (for recent related studies see e.g. \cite{inverseco}), we include, via detailed Monte Carlo simulations, the production of gamma rays in the annihilation event both via final state radiation and neutral pion decay. The black dashed line corresponds to the profile of Eq.~(\ref{eq:vl2}), while the solid line to a more conservative choice of a cored density profile slope in the innermost regions of the Galaxy, resulting in a normalization smaller by a factor $\sim 20$.

Both in the pure $e^\pm$ model (Fig.~\ref{fig:mod1}) and in the lepto-philic models (Fig.~\ref{fig:mod2}) a dark matter interpretation that gives a reasonable fit to both the PAMELA and the Fermi data is possible. The preferred range for the dark matter mass lies between 400 GeV and 1-2 TeV, with larger masses increasingly constrained by the H.E.S.S. results \cite{hess}. The required annihilation rates, when employing the dark matter density profile of Eq.~(\ref{eq:vl2}), imply typical boost factors ranging between 20 and 100, when compared to the value $\langle\sigma v\rangle\sim3\times 10^{-26}\ {\rm cm}^3/{\rm sec}$ expected for a thermally produced dark matter particle relic. While the detection of a signal from the Galactic center is not automatically implied, the expected flux in gamma-rays is likely sizable, but depends crucially on what is assumed for the inner slope of the dark matter density profile. All these considerations apply quite uniformly to the entire range of diffusion parameters and models we tested.

The outline of regions giving good fits to the PAMELA and to the Fermi data is squeezed to very large values in the case of dark matter pair-annihilating into gauge bosons, Fig.~\ref{fig:mod3}. The preferred mass range lies between 7-8 TeV (where a reasonable fit to the Fermi data is possible) and a few tens of TeV (where PAMELA data can also be explained). The required boost factor is very large, of the order of $10^4$. In this scenario, however, the H.E.S.S. data provide rather stringent constraints \cite{hess}.

\begin{figure}[!t]
 \centering 
 \includegraphics[width=4.5in]{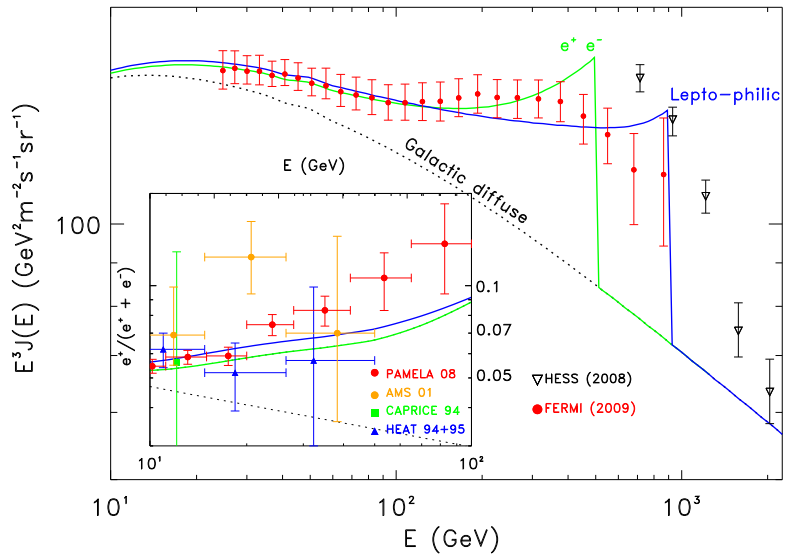}
 \caption{\footnotesize  \it Predictions for the CRE spectrum from two specific dark matter models, compared to current measurements.
 The same large-scale Galactic CRE components (dotted line) as in Fig.s \ref{fig:elepos_Monogem_242reac} and \ref{fig:elepos_random_pulsars}
 (model 0 in Tab. \ref{table1}) is used here. Note that the theoretical model curves showed in this plot do not account for the smearing due the finite experimental energy resolution.
 }
 \label{fig:fermi_DM}
\end{figure}

We also point out that such large values of the pair-annihilation rate are generically in contrast with the synthesis of light elements in the early Universe. In particular, the orange curve in the middle panel of Fig.~\ref{fig:mod3} shows the estimate of (Jedamzik 2004 \cite{Jedamzik:2004ip}) for the constraint from the over-production of the isotope ${}^6$Li from dark matter annihilation during Big Bang Nucleosynthesis. In addition, annihilation in gauge bosons produces an anomalous background of energetic neutrinos, which exceeds the current constraints (Yuksel et al. 2007 \cite{Yuksel:2007ac}) in almost all the parameter space compatible with $e^\pm$ data. In this respect, the gauge boson annihilation mode appears to be disfavored with respect to the previous two scenarios outlined above.
Other constraints on the dark matter pair annihilation rate from BBN include e.g. limits from measurements of the He$^3$/D ratio.
Ref.~\cite{HKKNM} recently showed that for the $W^+W^-$ annihilation final state, the He$^3$/D ratio gives even more stringent constraints than those
from $^6$Li we quote here and show in Fig.~\ref{fig:mod3}. The results of \cite{HKKNM} also indicate that for the $e^+e^-$ state and for lepto-philic models
bounds from BBN on the dark matter annihilation rate are weaker than those from the Fermi-LAT data (grey shaded regions in our figures).

For illustrative purposes, we select two reference choices for the mass and pair annihilation rate for a model annihilating into $e^+e^-$ (fig.~\ref{fig:mod1}) and for a ``lepto-philic'' case (fig.~\ref{fig:mod2}). 
We quote in Tab.~\ref{tab:DMMODELS} the parameter values for the models we employ.
We show the resulting $e^\pm$ spectra, summed with the conventional background we adopt in the present analysis, in Fig.~\ref{fig:fermi_DM}. 
In the insert of the same figure we also show the resulting positron fraction.

\begin{table}[ht]
\centering
\caption{Dark Matter parameters for the models shown in Fig.~\ref{fig:mod1}-\ref{fig:mod3}}
\begin{center}
\begin{tabular}{|c|c|c|c|c|c|c|}
\hline
\hline
Model & Ann. Final State & Mass (GeV) & $\langle\sigma v\rangle$
(cm$^3/$s) \cr
\hline
$e^+e^-$ & $e^+e^-$ & 500 & $9\times 10^{-25}$ \cr
Leptophilic & 33\%($e^+e^-$)+33\%($\mu^+\mu^-$)+33\%($\tau^+\tau^-$) &
900 & $4.3\times 10^{-24}$ \cr
\hline
\end{tabular}
\end{center}
\label{tab:DMMODELS}
\end{table}%

In summary, Fermi-LAT data on $e^\pm$ set constraints and provide information on the nature of particle dark matter models in relation to the production of energetic leptons in annihilation events in the Galaxy. Assuming an exotic origin for the data reported in (Abdo et al. 2009~\cite{Fermi_CRE_1}), we showed that the required dark matter setup is consistent with the PAMELA data and with the H.E.S.S. measurements. Specifically, we argued that models where dark matter only pair annihilates into charged leptons can give a satisfactory fit to the data for dark matter particle masses between 0.4 and 2 TeV, and for boost factors on the order of $10^2$.

\section{Discussion}\label{sec:discussion}

A common prediction of pulsar and DM interpretations of the combined Fermi-LAT and PAMELA data which we proposed 
 in Sec.\ref{sec:pulsars} and Sec.\ref{sec:DM} is a steadily growing positron fraction up to around 500 GeV.
AMS-02 which is planned for operation on the International Space Station in 2010, should be able to confirm/disprove this prediction in a few years (see e.g.  Rosier-Lees 2007~\cite{ams02}). Other experiments currently in the R\&D phase, such as PEBS (Gast et al. 2009 \cite{Gast:2009yh}) could also contribute to greatly improve our knowledge of the positron-to-electron fraction.
The question thus naturally arises as to whether it is possible to distinguish between a pulsar and a dark matter interpretation. 
While the positron ratio measurements alone are likely insufficient to settle the issue,  the shape of the high energy part of the electron spectrum might in principle provide a valuable tool (see e.g. Hall \& Hooper 2008 \cite{Hall:2008qu}, Pohl 2009 \cite{Pohl:2008gm}).  Generally, the cutoff expected in the electron spectrum in the DM scenario
(especially for the Kaluza-Klein DM models) is sharper than in the case of pulsars. The actual feasibility of discriminating between those interpretations from the experimental 
data, however, clearly depends on the details of both pulsars and particle dark matter models. It should also be noted that the theoretical model diagrams represented in this paper do not account for the instrumental response of the Fermi-LAT and H.E.S.S. experiments so that the differences among the signals expected in the pulsar and DM scenarios may be actually smaller than it appears from those figures.   Furthermore other Dark Matter models (see e.g. Bergstrom et al. 2009 \cite{Bergstrominprep} and Meade et al.
\cite{Meade:2009iu}), which have not discussed in this paper, have been recently showed to allow even better fits of Fermi-LAT, H.E.S.S. and PAMELA data.
Therefore, while the figures showed in this paper seem to favor a pulsar interpretation of  the Fermi-LAT measurements, a dedicated analysis, as well as more events at high energy, are needed to possibly settle this issue.  

It is appropriate mentioning here that in a recent paper (Malyshev et al. 2009 \cite{Malyshev:2009tw})  the authors claim that the possible observation of a smooth electron spectrum by Fermi-LAT would favour the DM interpretation of those measurements.  In a few words, their argument is that the contribution of multiple pulsars to the electron spectrum would result in significant fluctuations of the total electron + positron spectrum above few 100 GeV. That conclusion, however, was based on a choice of relevant parameters to match the spectrum measured by ATIC at around 600 GeV which is not confirmed by Fermi-LAT. In particular Malyshev et al. 
(2009~ \cite{Malyshev:2009tw}) adopt very high pulsar spin-down luminosities.  Furthermore, in that paper the cutoff energy in the pulsar injection spectrum $E_{\rm cut}$ is considerably higher than that assumed here.  In that case,  since $E_{\rm cut} > E_{\rm max}$,  a much sharper termination of the propagated spectrum from each contributing pulsar is expected (see the related discussion in Sec. \ref{sec:pulsar}).  However, as we argued in Sec.\ref{sec:pulsar} and  noted by Malyshev et al. themselves in their paper,
this is not  a necessary assumption for mature pulsars. It should be further noted that in this work we assumed an instantaneous $e^\pm$ release from pulsars. Adopting a more realistic finite duration may only smooth further the propagated electron spectrum. 

In principle, a possible {\it smoking gun} signature for the pulsar interpretation scenario might be provided by the IC emission in the direction of one of the closest mature pulsars. 
As the emission is expected to decrease with pulsar age, the region around Monogem is the most promising direction to look at.  As the electron spectrum is expected to be peaked  
at  $\sim  500~\GeV$ the IC scattering emission onto CMB photons will be maximal at around $1~\GeV$, hence both in EGRET and Fermi gamma-ray energy sensitivity range. 
We estimated the expected flux at that energy by integrating the electron spectral density, as given in Eq. \ref{eq:pulsar_spectrum},  divided by the IC time loss along Monogem line of sight (see e.g. Aharonian et al. 1997). We  found it to be about more than two orders of magnitude smaller than the diffuse gamma-ray flux observed by EGRET. It is therefore unlikely that this channel may allow a positive signature/disproof of the scenario proposed in Sec.\ref{sec:pulsars}. 

\begin{figure}[!t]
 \centering
  \includegraphics[width=4.5in]{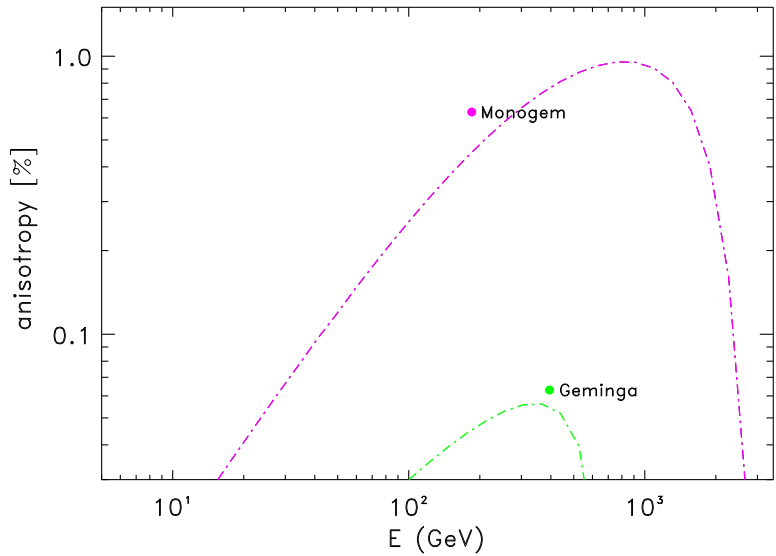}
 \caption{\it\small In this figure we show the electron + positron  expected anisotropy in the directions of Monogem and Geminga as a function of the energy
 under the same conditions adopted in Fig. \ref{fig:elepos_Monogem_242reac}. }
\label{fig:anisotropy}
\end{figure}

It has been suggested that a more promising approach may be offered by anisotropy measurements of the high energy CRE flux. As observed by several authors 
(see e.g. Kobayashi et al. 2004 ~\cite{Kobayashi:2003kp}, Hooper et al.  2008 \cite{Hooper:2008kg}, 
B\"ushing et al. 2008b ~\cite{Buesching:2008hr}) the electron emission from a few 100 pc distant pulsar may give rise to an observable anisotropy in the direction of that source. 
In the DM scenario a possible anisotropy is also expected pointing in the direction of the Galactic Center (note that Monogem and Geminga angular positions are very close and
almost opposite to the GC) or of local DM clumps. The latter, however,  will be unlikely in the same direction of a nearby pulsar.   
Here we estimate the anisotropy induced by the most luminous nearby pulsars, namely Monogem and Geminga, under the same hypothesis which has been used in Sec. 3.2. 
From eq. \ref{eq:pulsar_spectrum} we find  
\begin{equation}
{\rm Anisotropy}  =  \frac{3}{2 c} \frac{r}{t-t_0}~\left( \frac{1 - (1 - E/E_{max}(t))^{1 - \delta} } {(1 - \delta)E/E_{max}(t) } \right)^{-1} \; \frac{N_e^{\rm PSR}(E)}{N_e^{\rm tot}(E)}
\end{equation}
where $N_e^{\rm PSR}$ and $N_e^{\rm tot}$ are the electron spectra from the pulsar and its sum to the large scale Galactic plus distant pulsar components. 
We found that (due to its old age) the contribution from Geminga is negligible so that the anisotropy should be dominated by Monogem electrons (see Fig. \ref{fig:anisotropy}). 
   
This anisotropy is comparable with that estimated in previous publications on the basis of PAMELA results (see e.g.  Buesching et al. 2008 \cite{Buesching:2008hr}, 
Hooper et al. 2009 \cite{Hooper:2008kg}) and should  be observable by Fermi-LAT in a few years of data taking (a dedicated analysis to determine the CRE anisotropy 
Fermi-LAT sensitivity is beyond the scopes of this work). 

\section{Conclusions}\label{sec:conclusions}

We report on some possible interpretations for the cosmic ray electron-plus-positron (CRE) spectrum measured by Fermi-LAT. The measured CRE flux is significantly harder than previously believed, and it does not show any sharp feature in the multi-hundred GeV range,  although there are hints of an extra-component between 100 and 1000 GeV. 

In the context of astrophysical interpretations to the CRE data, we discussed in the present analysis the case of a single large-scale diffuse Galactic (GCRE) component, and a two-component scenario which adds to the GCRE flux a primary component produced by mature pulsars.
In the GCRE scenario, a spatially continuous distribution of primary CRE sources in the Galactic disk, provides a satisfactory explanation to the Fermi-LAT CRE data for several 
combinations of the injection spectral index $\gamma_0$ and the CR propagation parameters. In particular we showed this to be the case for $\gamma_0 = 2.42$ (for a CR 
propagation set-up with a power-law spectral index of the diffusion coefficient $\delta = 1/3$) and for $\gamma = 2.33$ (for $\delta = 0.6$). We verified that at least the former 
scenario is compatible with current preliminary data on the diffuse gamma-ray radiation measured by Fermi at intermediate Galactic latitudes.
Although the GCRE scenario requires a spectral break to be consistent with the H.E.S.S. data at energies larger than a TeV, we showed that it may arise as a consequence of the
stochastic spatial and temporal distribution of nearby sources. This scenario, however, is in sharp tension with the PAMELA data on the positron fraction, more than previously 
considered in the framework of GCRE models, as a consequence of the hardness of the electron plus positron spectrum measured by Fermi-LAT.  
Furthermore,  a tension may also be present between these GCRE models fitting the Fermi-LAT CRE spectrum and pre-Fermi experimental data below 10 GeV.

Taking into account nearby mature pulsars as additional sources of high-energy CRE, we showed that both the PAMELA positron excess and the Fermi-LAT CRE data are naturally explained by nearby (distances less than one kpc) known objects. We also considered the overall effect of a combination of all known pulsars within a larger distance, and varied the parameters that affect the pulsars' CRE injection spectrum, and concluded that the observed CRE and positron abundance spectral features are all consistently reproduced.

We also considered another possible primary source of high-energy CRE: the annihilation or decay of particle dark matter in the Galactic halo. Fermi-LAT CRE data do not confirm the sharp spectral feature in the 500-1000 GeV range that prompted several studies to consider a dark matter particle mass in that same range. Yet, we showed that a dark matter particle annihilating or decaying dominantly in leptonic channels, and with a mass between 400 GeV and 2 TeV is compatible with both the positron excess reported by PAMELA and with the CRE spectrum measured by Fermi-LAT.  It is understood that the DM models considered in this paper are only a representative and limited subset of a much wider collection.  

While we found that the pulsar interpretation seems to be favored by Fermi-LAT CRE data, a clear discrimination between this and the dark matter scenario is not 
possible on the basis of the currently available data and requires to consider complementary observations. 

Other possible scenarios, as that recently proposed by Blasi (2009~\cite{Blasi:2009hv}), which have not been discussed in this paper, may also offer viable interpretations of Fermi-LAT CRE data.

Exciting times lie ahead towards the understanding of the nature of high-energy CRE: future Fermi-LAT data will (i) extend the energy range both to lower and to higher energies than reported so far, (ii) allow anisotropy studies of the arrival direction of high-energy CRE, which could conclusively point towards one (or more than one) nearby mature pulsar as the origin of high-energy CRE, and (iii) deepen our understanding of pulsars via gamma-ray observations, and via the discovery of new gamma-ray pulsars, potentially extremely relevant as high-energy CRE sources. 
 Last but not least, Fermi measurements of the spectrum and angular distribution of the diffuse gamma-ray emission of the Galaxy will also shed light on the nature and spatial distribution of CRE sources. 
\section*{Acknowledgments}

The Fermi LAT Collaboration acknowledges generous ongoing support from a number of agencies and institutes that have supported both the development and the operation of the LAT as well as scientific data analysis. These include the National Aeronautics and Space Administration and the Department of Energy in the United States, the Commissariat ˆ l'Energie Atomique and the Centre National de la Recherche Scientifique / Institut National de Physique NuclŽaire et de Physique des Particules in France, the Agenzia Spaziale Italiana and the Istituto Nazionale di Fisica Nucleare in Italy, the Ministry of Education, Culture, Sports, Science and Technology (MEXT), High Energy Accelerator Research Organization (KEK) and Japan Aerospace Exploration Agency (JAXA) in Japan, and the K. A. Wallenberg Foundation, the Swedish Research Council and the Swedish National Space Board in Sweden.

Additional support for science analysis during the operations phase from the following agency is also gratefully acknowledged: the Istituto Nazionale di Astrofisica in Italy. 

D.G. is supported by the Italian Space Agency under the contract AMS-02.ASI/AMS-02 n.I/035/07/0.  
He also thanks the Dipartimento di Fisica dell'Universit\`a di Padova and UniverseNet EU Network under contract n. MRTN-CT-2006-035863 for partial finantial support.
S.P. is partly supported by US DoE Contract DEFG02-04ER41268 and by NSF Grant PHY-0757911.
E.D.B. and T.K. are supported by US DoE Contract DE-AC02-76-SF00515.

We thank the anonymous referee for several useful comments.

 \renewcommand{\theequation}{A-\arabic{equation}}
 \setcounter{equation}{0}  
 \section*{APPENDIX}  

The spectrum of electrons and positrons from a point source can be readily calculated by solving the transport equation
\begin{equation}
\label{eq:transport} \frac{\partial N_e(E,t,\vec{r})}{\partial t}
- D(E) \nabla^2 N_e - \frac{\partial}{\partial E} (b(E) N_e) =
Q(E,t,\vec{r})
\end{equation}
where $N_e(E,t,\vec{r})$ is the number density of  $e^\pm$ per unit energy,  $D(E)$ is the diffusion coefficient (assumed to be spatially uniform), $b(E)$ 
the rate of energy loss and $Q(E,t,\vec{r})$ the source term. Here we neglect convection and re-acceleration, their role being negligible above $\sim 10$ GeV  
especially on short  $\sim 100~\pc$ distances.  The diffusion coefficient is assumed to have the usual power law dependence on energy $D(E) = D_0~(E/E_0)^\delta$. 
Both normalization and the power law index are chosen to be the same as adopted in GALPROP to model the GCRE component.

If we consider a bursting-like source with a general energy spectrum $Q(E,t,\vec{r}) = Q(E)~\delta(t-t_0)~\delta(\vec{r})$, where $t_0$ is the injection time (the instant in which
the particles are released from the source into the ISM), and $\vec{r}$ is the distance to the source, the solution of Eq. (\ref{eq:transport}) is (Ginzburg \& Putskin 1976 
\cite{Ginzburg},  Atoyan et al. 1995 \cite{Atoyan:95}):
\begin{equation}
\label{eq:green_function} 
N_e(E,~ t,~\vec{r}) = \frac{Q(E_i)~b(E_i)}{\pi^{3/2}~b(E)~r_{\rm diff}^3(E,t)}~e^{- \left(r/r_{\rm diff}(E,t) \right)^2 }
\end{equation}
where $E_i$ is the initial energy of particles which are cooled down to energy $E$ during time $t-t_0$, and $r_{\rm diff}$ is the
diffusion distance (i.e. the propagation distance over which the electron lose half of its energy).

In our case, the source term is taken as
\begin{equation}
\label{eq:inj_spectrum}
Q(E,t,\vec{r}) = Q_0~\left(\frac{E}{1\, {\rm GeV}}\right)^{- \Gamma}~e^{(- E/E_{\rm cut})}~\delta(t-t_0)~\delta(\vec{r})  
\end{equation}
and the energy loss rate $b(E)$, since only syncrotron and IC losses are relevant, is expressed as $b(E) = b_0 ~ E^2$, where $b_0 = 1.4 \times 10^{-16}~
\GeV^{-1} \s^{-1}$  which is taken to be same as in GALPROP at the Sun position.
In this case the solution is:
\begin{equation}
\label{eq:pulsar_spectrum} 
N_e(E,~ t,~\vec{r}) =
\frac{Q_0}{\pi^{3/2}~r_{\rm diff}^3}~\left( 1 - E/E_{max}
\right)^{\Gamma -2}~\left(\frac{E}{1\, {\rm GeV}}\right)^{- \Gamma}~e^{- \fraction{E}{\left(1 -
E/E_{max}\right)~E_{\rm cut}}}~ \ e^{ - \left(r/r_{\rm diff}(E)
\right)^2 }
\end{equation}
for $E < E_{max}$, and $0$ otherwise, where the diffusion distance is given by
\begin{equation}
r_{\rm diff}(E,t) \,\approx\, 2 ~ \sqrt{D(E) (t-t_0) ~\frac{1 - (1 -
E/E_{max}(t))^{1 - \delta}}{(1 - \delta)E/E_{max}(t) } }
\end{equation}
and
\begin{equation}
\label{eq:Emax}
E_{\max}(t) \,=\, \frac{1}{b_0 ~ (t-t_0) }  
\end{equation}
It should be noted that sources injecting electrons at a time $t_0$ with $t - t_0 \ll \tau_{\rm diff} \simeq r^2 / D(E)$ cannot contribute to the
electron flux reaching the observer.


\end{document}